\documentclass[aps,prl,showpacs,twocolumn,amsmath,amssymb,superscriptaddress]{revtex4-1}

\usepackage{tabularx}
\usepackage{bm}
\usepackage{graphicx}

\usepackage{color}

\usepackage{multirow}
\usepackage{dcolumn}
\usepackage{amssymb,amscd,xypic,bm,wasysym}
\usepackage{float}
\usepackage{cleveref}
\newcolumntype{Y}{>{\centering\arraybackslash}X}

\usepackage{soul}  

\begin{document}

\title{Manipulating Anomalous Hall Antiferromagnets with Magnetic Fields}

\author{Hua Chen}
\affiliation{Department of Physics, Colorado State University, Fort Collins, CO 80523, USA}
\affiliation{School of Advanced Materials Discovery, Colorado State University, Fort Collins, CO 80523, USA}
\author{Tzu-Cheng Wang}
\affiliation{Department of Physics and Center for Theoretical Physics, National Taiwan University, Taipei 10617, Taiwan}
\author{Di Xiao}
\affiliation{Department of Physics, Carnegie Mellon University, Pittsburgh, PA 15213, USA}
\author{Guang-Yu Guo}
\affiliation{Department of Physics and Center for Theoretical Physics, National Taiwan University, Taipei 10617, Taiwan}
\affiliation{Physics Division, National Center for Theoretical Sciences, Hsinchu 30013, Taiwan}
\author{Qian Niu}
\affiliation{Department of Physics, the University of Texas at Austin, Austin, TX 78712, USA}
\author{Allan H. MacDonald}
\affiliation{Department of Physics, the University of Texas at Austin, Austin, TX 78712, USA}
    
\begin{abstract}
The symmetry considerations that imply a non-zero anomalous Hall effect (AHE) in certain non-collinear antiferromagnets also imply both non-zero orbital magnetization and a net spin magnetization. We have explicitly evaluated the orbital magnetizations of several anomalous Hall effect antiferromagnets and find that they tend to dominate over spin magnetizations, especially so when spin-orbit interactions are weak. Because of the greater relative importance of orbital magnetization the coupling between magnetic order and an external magnetic field is unusual. We explain how magnetic fields can be used to manipulate magnetic configurations in these systems, pointing in particular to the important role played by the response of orbital magnetization to the Zeeman-like spin exchange fields.
\end{abstract}

\maketitle

{\it Introduction}---We have previously~\cite{chen_2014} pointed out that spin-orbit interactions induce an anomalous Hall conductivity {\it i.e.} an antisymmetric contribution to the conductivity tensor $\sigma_{\alpha\beta} = \partial j_\alpha/\partial E_\beta$, in some common antiferromagnets (AFMs) with non-collinear magnetic order. Because the anomalous Hall effect (AHE) is usually associated with ferromagnetism, we refer to these systems as AHE AFMs. One way to understand the finite anomalous Hall conductivity of AHE AFMs is to view it as a time-reversal-odd pseudovector $\sigma^{\rm AH}_\alpha = \epsilon_{\alpha\beta\gamma} \sigma_{\beta\gamma}/2$ that only vanishes in magnetic systems when required to do so by some lattice symmetry. This idea of spatial symmetry controlled AHE has also been extended to collinear AFMs~\cite{sivadas_2016}.

Since the total magnetization is also a time-reversal-odd pseudovector, it must be nonzero in AHE AFMs. Indeed, Mn$_3$Ir, the prototypical AHE AFM identified in Ref.~\cite{chen_2014}, has a finite magnetization~\cite{szunyogh_2009,leblanc_2013}, as do other AHE AFMs such as Mn$_3$Sn and Mn$_3$Ge \cite{kubler_2014, nakatsuji_2015, nayak_2016, kiyohara_2016}. It is precisely because of the nonzero magnetization, that the sign of the AHE can be flipped by reversing the magnetic field direction in experiments. However, the microscopic picture of magnetization in AHE AFMs is far from clear. In particular, it is expected that typical AHE AFMs should have vanishingly small total spin magnetization due to the much larger exchange coupling than the magnetic anisotropy of sublattice moments. As a result, the orbital contribution to the total magnetization \cite{hirst_1997, shindou_2001, xiao_2005, thonhauser_2005, ceresoli_2006, shi_2007} is no longer negligible, and could play a key role in determining how AHE AFMs respond to external magnetic fields.  

Our goal in this Letter is to develop a quantitative description of manipulating the order parameter direction of AHE AFMs coherently using orbital magnetic fields, which is appropriate for those AHE AFMs with dominating orbital magnetization over the spin contribution. To this end, we first provide a general criterion, backed by first-principles calculations, for searching for such orbital-magnetization-dominant AHE AFMs. We then point out, in the framework of relativistic spin density functional theory (SDFT), that the orbital magnetic field reorients the order parameter through an unusual orbital-spin susceptibility, for which we give a convenient formula based on linear response theory. With these preparations, we finally explain our method for investigating field-induced coherent order parameter switching in such AHE AFMs, by keeping track of energy extrema evolution in the configuration space, and illustrate the various unusual switching behaviors by applying this approach to a toy model mimicking Mn$_3$Ir.

{\it Ground State Orbital and Spin Magnetizations}---Orbital magnetization arises from circulating electron currents. In a finite system it can be unambiguously defined as the expectation value of $-\frac{1}{2}{\bf j} \times {\bf r}$~\cite{hirst_1997}. In an extended system this definition of orbital magnetization becomes ambiguous because the position operator is unbounded. Historically this conundrum posed both conceptual and practical challenges, but have been fully solved recently \cite{xiao_2005, thonhauser_2005, ceresoli_2006, shi_2007}. In particular, we now know that there are two gauge-invariant contributions to the total orbital magnetization of an extended system, due to the magnetic moments of individual Bloch wave packets and to the Berry phase modification of the electron density of states in a magnetic field, respectively~\cite{xiao_2005, souza_2008}.

\begin{table}
	\caption{Ground state spin and orbital magnetization (in $m\mu_B$ per formula unit) for some common AHE AFMs. The partial orbital magnetizations $M_{\rm orb}^1$ and $M_{\rm orb}^2$ are respectively the Bloch state orbital moment and magnetic-field-dependent density-of-states contributions.}\label{tab:morb}
	\centering
	{\def\arraystretch{1.2}
	\begin{tabularx}{0.45\textwidth}{c *{5}{Y}}
		\\
		\hline\hline
		   &  $M_{\rm spin}$    &  $M_{\rm orb}^1$  & $M_{\rm orb}^2$    & $M_{\rm orb}^{\rm tot}$ \\\hline
		Mn$_3$Ir         & 26.9  & -76.7  & 106.1  &  29.7 \\
		Mn$_3$Pt         & 11.2  & -17.0  &  29.4  &  12.2 \\
		Mn$_3$Rh         & 2.4  &  -24.0  &  35.0  &  11.0 \\	         
		Mn$_3$Sn         & 0.9  &   40.5  & -42.5  &  -2.0 \\
		Mn$_3$Ge         & 0.9  &  -17.5  &  35.2  &  17.7 \\\hline\hline
	\end{tabularx}
    }
\end{table}

To verify that orbital magnetization has a larger relative importance in AHE AFMs we have calculated both orbital and spin magnetizations in Mn$_3$Ir, Mn$_3$Pt, Mn$_3$Rh, Mn$_3$Sn, and Mn$_3$Ge, all AHE AFMs according to previous work \cite{chen_2014, kubler_2014, nakatsuji_2015, nayak_2016, kiyohara_2016}, listed in Table~\ref{tab:morb}. The orbital magnetization $M_{\rm orb}$ is calculated with the zero-temperature expression given in e.g.~\cite{shi_2007} using Wannier interpolation of results from relativistic SDFT~\cite{supp}, which adds corrections from spin-orbit coupling to the Kohn-Sham single particle equations, but employs exchange-correlation energy functionals that retain the structure of the non-relativistic limit~\cite{MacDonald_1979}. We find that $M_{\rm orb}$ is at least comparable to the total spin magnetization $M_{\rm spin}$ in size, and that it is much larger than the latter in certain materials, {\it e.g.} Mn$_3$Rh. This is in sharp contrast to conventional metallic ferromagnets such as Fe in which orbital magnetization is more than one order of magnitude smaller than spin magnetization. We are also aware of earlier SDFT calculations showing the importance of orbital magnetization in Mn$_3$Sn \cite{sandratskii_1996} prior to the establishment of a gauge-invariant form of the orbital magnetization in crystalline solids.
 
Interestingly, comparing $M_{\rm orb}$ and $M_{\rm spin}$ across Table \ref{tab:morb}, we see that heavier elements have smaller $M_{\rm orb}/M_{\rm spin}$ values. This trend can be understood by taking spin-orbit coupling as a weak perturbation \cite{Richter_1986, Bruno_1989, bruno_1993}, as we explain below. We consider first the atomic limit in which spin-orbit coupling can be approximated by $\lambda_{\rm so} {\bf L}\cdot {\bf S}$. Here $\bf L$ and $\bf S$ are the orbital and spin angular momentum operators that are proportional with appropriate $g$-factors to the local orbital and spin magnetic moments. It follows that magnetic order, which leads to a nonzero spin density averaged over an atomic sphere surrounding each magnetic atom, results in an effective magnetic field that couples directly to the local orbital moment. We write this effective coupling as $-{\bf M}_{\rm orb}\cdot {\bf H}$, where ${\bf M}_{\rm orb}=- g_o\mu_{B} {\bf L}/\hbar$ and ${\bf H} =  \hbar \lambda_{\rm so} S \hat{\Omega}/{g_o\mu_B}$, with $S$ and $\hat{\Omega}$ the magnitude and the direction of the local spin density, and $g_o$ the appropriate g-factor. The orbital magnetization is then the orbital-orbital susceptibility $\overleftrightarrow{\chi}_{\rm o}$, a rank-2 tensor that is non-zero even in the absence of spin-orbit coupling, times this effective magnetic field. It follows that the orbital magnetization is linear in spin-orbit coupling strength in the perturbative limit. In noncollinear antiferromagnets, the total orbital magnetization is a sum over sublattices of orbital-orbital susceptibilities times local orbital magnetic fields \cite{supp}.

In a similar way, the spin-canting that produces a non-zero total spin magnetization in these AHE AFMs can be viewed as the net spin density induced by the orbital magnetic field through a \emph{spin-orbital susceptibility} $\overleftrightarrow\chi_{\rm so}$ that connects spins and orbital magnetic fields. Since $\overleftrightarrow\chi_{\rm so}$ is clearly zero in the absence of spin-orbit coupling, it must be at least linear in $\lambda_{\rm so}$, and the spin-canting must therefore be at least of 2nd order. The same conclusion can be reached by relating $\overleftrightarrow\chi_{\rm so}$ to magnetocrystallline anisotropy \cite{supp}. It will be shown below that $\overleftrightarrow\chi_{\rm so}$ plays a central role in the reorientation of the noncollinear magnetic order parameters by external magnetic fields. 

\begin{figure}
	\begin{center}
		\includegraphics[width=2.6 in]{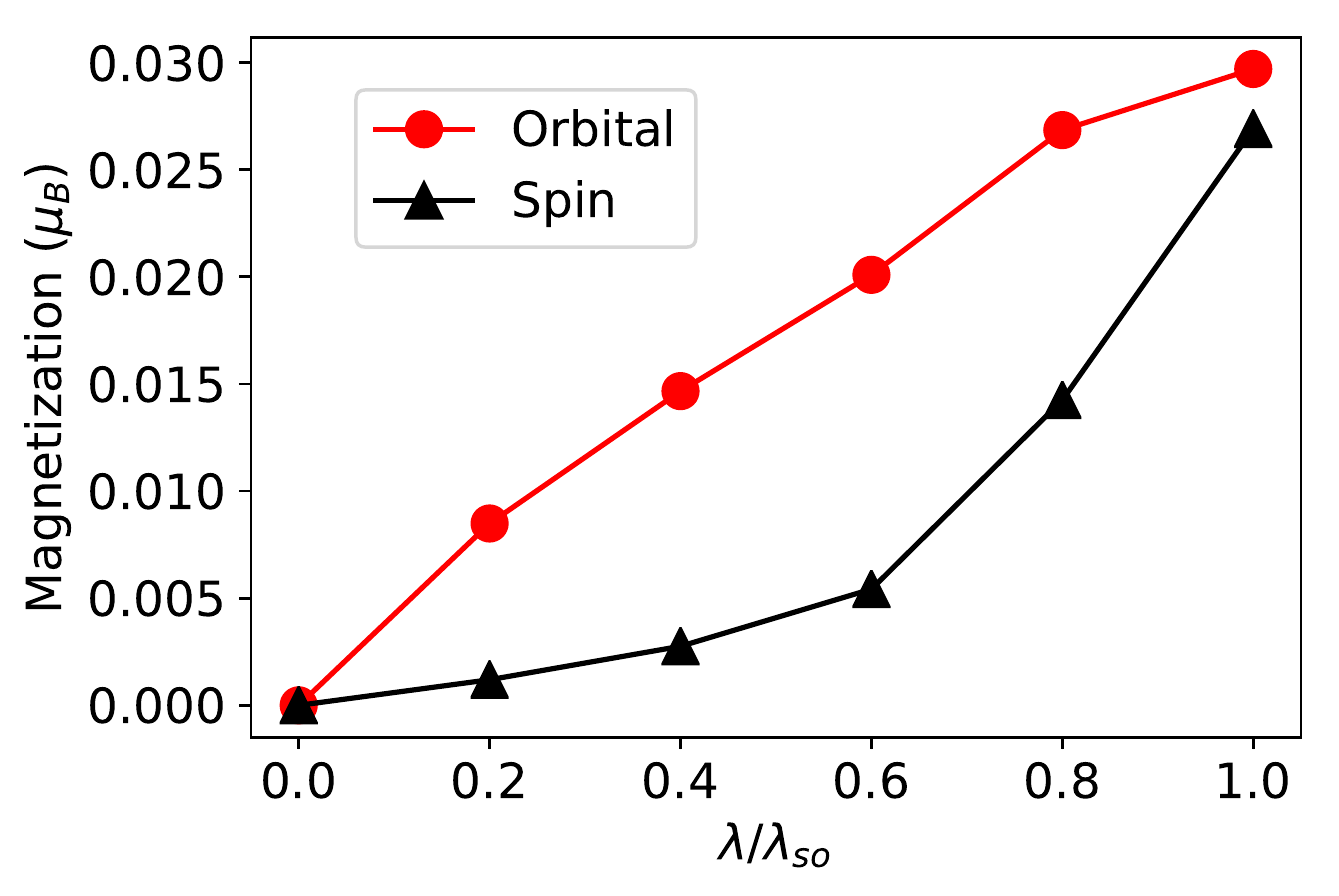}
	\end{center}
	\caption{Dependence of net $M_{\rm spin}$ and $M_{\rm orb}$ on spin-orbit coupling strength in Mn$_3$Ir. $\lambda/\lambda_{\rm so}$ is the ratio of spin-orbit coupling strength to its realistic value.}
	\label{fig:mvslambda}
\end{figure}

Although these atomic limit considerations do not strictly apply to metallic AFMs, we expect that the general trend should still hold. As an explicit check, we calculated the total orbital and spin magnetizations of Mn$_3$Ir {\it vs.} spin-orbit coupling strength by artificially varying the speed of light when generating the fully-relativistic pseudopotentials. The results shown in Fig.~\ref{fig:mvslambda} agree well with the qualitative picture explained above. It follows that in an AHE AFM family of given symmetries, larger $M_{\rm orb}/M_{\rm spin}$ values should be expected in materials with weaker, not stronger, atomic spin-orbit coupling.

{\it Manipulating AHE AFM Order with an Orbital Magnetic Field}---Having established the importance of orbital magnetization in AHE AFMs, below we discuss the order parameter reorientation induced by orbital magnetic fields within the relativistic SDFT formalism. Important differences between the present formulation and the conventional approach of solving the Landau-Lifshitz-Gilbert (LLG) equation for a classical spin model with local Zeeman coupling to external fields will be discussed at the end.

We first consider the simpler case of a ferromagnet in which the order parameter is a vector that specifies the spin-orientation $\hat{\Omega}$. Because the energy scales associated with external magnetic fields are small, it is sufficient to account only for the contribution to energy that is of first order in $\bf H$, namely the coupling of $\bf H$ to total magnetization. Minimizing total energy in the presence of a field then yields  
\begin{eqnarray} \label{eq:Hbalance}
0 =  \delta E_{\rm ani}(\delta\hat{\Omega}) - \delta{\bf M}(\delta\hat{\Omega})\cdot {\bf H},
\end{eqnarray}
where $E_{\rm ani}$ is the dependence of energy on order parameter direction in the absence of a field. When $\bf M$ is purely due to spin its magnitude is essentially fixed at the saturation magnetization $M_{\rm s}$. Eq.~\eqref{eq:Hbalance} then simply implies that the magnetization direction adjusts so that the anisotropy field ${\bf H}_{\rm ani} \equiv -\delta E_{\rm ani}/(M_{\rm s} \delta \hat{\Omega})$ cancels the external magnetic field. When $\bf M$ is dominated by the orbital contribution, on the other hand, Eq.~\eqref{eq:Hbalance} must be generalized to 
\begin{eqnarray} \label{eq:morbbalance}
{\bf H}_{\rm ani} + \frac{\delta {\bf M}_{\rm orb}}{M_{\rm s} \delta \hat{\Omega}} \cdot {\bf H} = 0.
\end{eqnarray}

To go further, we discuss the meaning of Eq.~\eqref{eq:morbbalance} within the framework of relativistic SDFT. For magnetic systems SDFT has the convenience of explicitly accounting for the Zeeman-like exchange coupling between the magnetic condensate and the Kohn-Sham quasiparticle spins in the exchange-correlation potential. Although the relativistic SDFT has some subtle disadvantages~\cite{Bruno_2001}, notably a failure~\cite{Galanakis_2005} to capture the interaction physics responsible for Hund's second rule, it is regularly and successfully applied and is built into common electronic structure software packages. Its practical success is likely due to the fact that the degree to which local spin alignment reduces interaction energies is not strongly altered by relativistic corrections. 

In this formalism $\hat{\Omega}$ enters the exchange-correlation potential in the form of $-\Delta_{\rm ex} \hat{\Omega}\cdot {\bf S} \equiv - g \mu_{\rm B} {\bf H}_{\rm spin} \cdot {\bf S}/\hbar$, where $\Delta_{\rm ex}$ is the exchange field strength. Using a simplified notation in which the variation of $\Delta_{\rm ex}$ within an atomic cell is left implicit, we have
\begin{eqnarray}
\frac{\delta {\bf M}_{\rm orb}}{M_{\rm s} \delta \hat{\Omega}} = \frac{\hbar\Delta_{\rm ex}}{g\mu_{\rm B} M_{\rm s}}\frac{\delta {\bf M}_{\rm orb}}{\delta {\bf H}_{\rm spin}} = \frac{\hbar\Delta_{\rm ex}}{g\mu_{\rm B} M_{\rm s}} \overleftrightarrow \chi_{\rm os},
\end{eqnarray} 
where $g\approx -2$ is the Lande g-factor, and $\overleftrightarrow\chi_{\rm os} = \overleftrightarrow\chi_{\rm so}^T$  is the \emph{orbital-spin susceptibility} discussed further below. With this notation Eq.~\eqref{eq:Hbalance} becomes
\begin{eqnarray}
{\bf H}_{\rm ani} = - \frac{\hbar\Delta_{\rm ex}}{g\mu_{\rm B} M_{\rm s}} \overleftrightarrow \chi_{\rm os}\cdot {\bf H}.
\end{eqnarray}
It follows that when the magnetization is orbitally dominated, the anisotropy field must be balanced by an adjustment in ${\bf M}_{\rm orb}$ produced by the orbital-spin susceptibility $\overleftrightarrow \chi_{\rm os}$ which, among the various magnetic susceptibility contributions identified in solid state systems \cite{misra_1972, fukuyama_1969, gao_2015, ogata_2015}, is the one seldom addressed in the literature~\cite{roth_1962, misra_1969, misra_1972}. 

$\overleftrightarrow \chi_{\rm os}$ in crystals can be calculated using the standard linear response theory which we sketch below. More details can be found in \cite{supp}. To apply a uniform orbital magnetic field to we consider a periodic vector potential ${\bf A}({\bf r}) = \frac{{\bf B}_{\rm orb}\times {\bf q}}{q^2}\sin({\bf q}\cdot{\bf r})$, then take the $q\rightarrow 0$ limit~\cite{fukuyama_1969, shi_2007} with ${\bf q}\cdot {\bf B}_{\rm orb} = 0$ \cite{note1}. One can then obtain for a grand canonical ensemble \cite{supp}
\begin{eqnarray}\label{eq:chisogf}
&&\chi_{\rm os}^{\alpha\beta} = -\frac{e\hbar g \mu_{\rm B}}{4} k_{\rm B} T \,   \epsilon_{\alpha\gamma\delta} \\\nonumber
&&\times {\rm Im}\sum_{n}\int[d\bm k] \,  {\rm tr}\left(G_0 v^{\gamma} G_0 v^\delta G_0 \sigma^\beta \right),
\end{eqnarray}
where Greek letters label Cartesian coordinates $x,y,z$, $G_0$ is the Kohn-Sham thermal Green's function, $\bf v$ is the velocity operator, $\bf \sigma$ is the spin-space Pauli matrix vector, $n$ is a fermionic Matsubara frequency label. Completing the Matsubara sum then yields a variety of terms that can be grouped as either Fermi-surface or Fermi-sea contributions \cite{supp}, which was not done in \cite{misra_1972}. The Green's function formalism is convenient when generalizing the theory to cover disorder and interaction effects \cite{fukuyama_1969, supp}.

We now turn to the specific case of AHE AFMs, in which it is convenient to view the magnetic sublattice dependent spin-density directions $\hat{\Omega}_i$ ($i$ labels the total $N$ magnetic sublattices) as the order parameter. Because the exchange coupling between local moments is strong, the relative orientations between local moments on different sublattices are normally nearly fixed. Then, as in the case of a classical rigid body, the number of parameters can be reduced to three for any $N$ \cite{andreev_1980, dombre_1989, gomonay_2015, ulloa_2016}. Generalization to include non-rigid rotation, which is necessary in, e.g., the case of Mn$_3$Sn, is discussed in \cite{supp}. The counterpart of Eq.~\eqref{eq:Hbalance} for the noncollinear case is 
\begin{eqnarray}\label{eq:Ebalance}
0 = \delta E_{\rm ani}(\delta{\bf \omega}) - \delta{\bf M}(\delta{\bf \omega})\cdot {\bf H},
\end{eqnarray}
where ${\bf \omega}$ represents the three variables parameterizing the three-dimensional rotation group SO(3). For infinitesimal rotations the three components of $\delta{\bf \omega}$ commute, and can be chosen as infinitesimal rotation angles around the three Cartesian axes $\delta \omega_\alpha$. It follows that 
\begin{eqnarray}\label{eq:nclbalance}
\frac{\delta E_{\rm ani}}{\delta \omega_\alpha} &=& \frac{\delta {\bf M}_{\rm orb}}{\delta \omega_\alpha} \cdot {\bf H} = {\bf H} \cdot \sum_{i=1}^N \frac{\delta {\bf M}_{\rm orb}}{\delta \hat{\Omega}_i} \cdot \frac{\delta \hat{\Omega}_i}{\delta \omega_\alpha} \\\nonumber
&=& \frac{\hbar\Delta_{\rm ex}}{g\mu_{\rm B}} H_\lambda \sum_{i=1}^N (\chi_{\rm os}^i)_{\lambda \gamma} \epsilon_{\gamma\alpha\beta} \, \Omega^i_{\beta}.
\end{eqnarray}  
where Greek letters label $x,y,z$, $\overleftrightarrow\chi_{\rm os}^i$ is the total orbital response to a local Zeeman field on sublattice $i$, which can be evaluated by using Eq.~\eqref{eq:chisogf} and projecting the spin operator onto site $i$. The Levi-Civita symbol comes from the antisymmetric infinitesimal rotation matrix in Cartesian coordinates.

With above preparations we propose the following strategy for studying coherent magnetic switching in AHE AFMs. Switching through domain nucleation and growth will be discussed elsewhere. With a microscopic Hamiltonian we can identify energy extrema that satisfy Eq.~\eqref{eq:nclbalance}. These correspond to local minima, maxima, and saddle points in the SO(3) parameter space. Both the positions in SO(3) space and the energies of these extrema change smoothly with increasing external magnetic field. Whenever a minimum is converted to a saddle point, magnetic switching to a new minimum can proceed. For numerical implementation one can discretize the SO(3) space, calculate $E_{\rm ani}$, ${\bf M}_{\rm orb}$, $\frac{\delta E_{\rm ani}}{\delta {\bf \omega}}$ and $\overleftrightarrow \chi_{\rm os}^i$ at each grid point, and search for the $\bf H$-dependent energy extrema.

\begin{figure}
	\begin{center}
		\includegraphics[width=3 in]{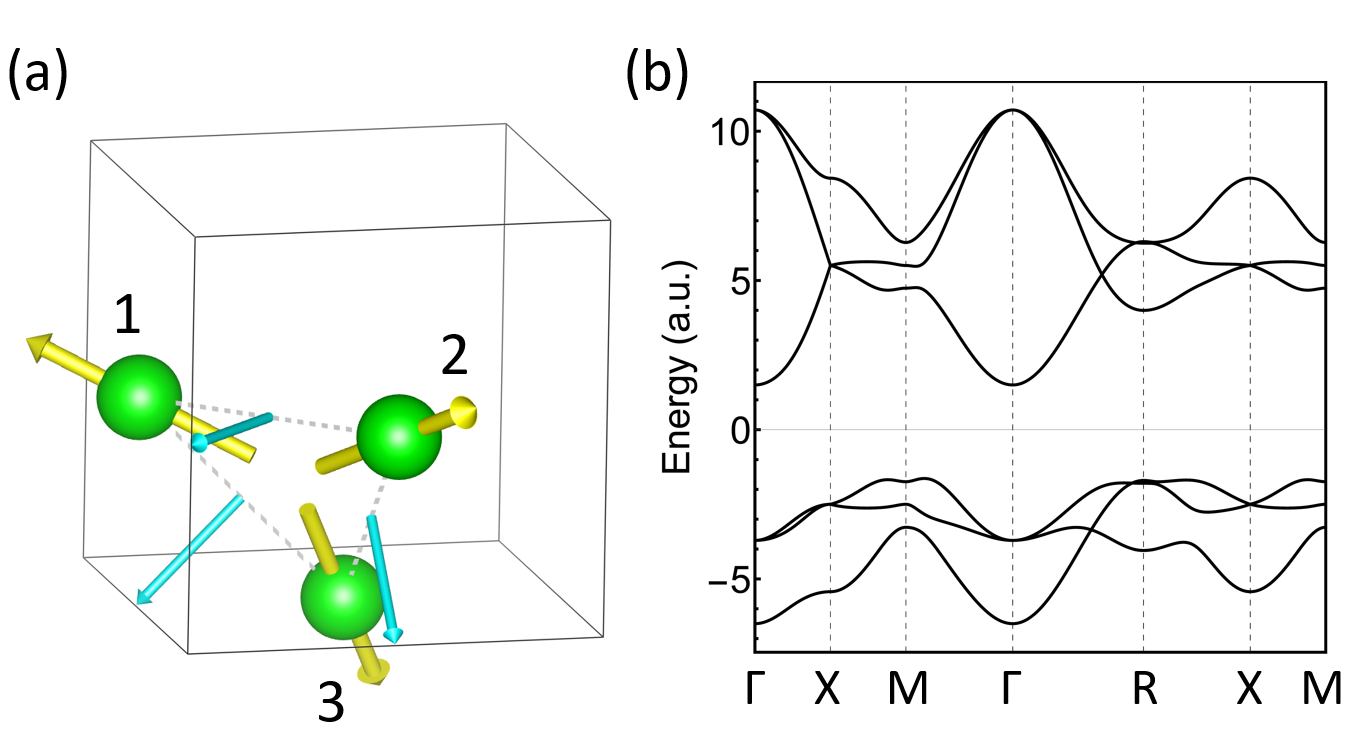}
	\end{center}
	\caption{(a) Structure of a $s$-$d$ model resembling Mn$_3$Ir with its bands shown in (b). The smaller arrows in (a) represent the $C_2$ axes $\hat{\eta}_{mn}$ in Eq.~\eqref{eq:Hso}. }\label{fig:1}
\end{figure}

{\it Application to a Model AHE AFM}---We now give an example of the procedure proposed above using a toy model that mimics the magnetic structure of Mn$_3$Ir. We consider a $1/4$-depleted fcc lattice (Fig.~\ref{fig:1}), with an $s$-orbital on each site, nearest-neighbor hopping, and sublattice-dependent exchange fields whose directions replicate the triangular antiferromagnetic order of Mn$_3$Ir. We add spin-orbit coupling $H_{\rm so}$, being careful to respect the $C_2$ symmetry axis $\hat{\eta}$ along bond-dependent lines (see Fig.~\ref{fig:1}) that pass through the center of each nearest neighbor bond: 
\begin{eqnarray}\label{eq:Hso}
H_{\rm so} = \sum_{\langle im,jn\rangle\alpha\beta}\text{i} t_{\rm so} ( 
\hat{d}_{im,jn} \times \hat{\eta}_{mn})\cdot{\bf \sigma}_{\alpha\beta} \; c_{im\alpha}^\dag c_{jn\beta}.
\end{eqnarray} 
Here $ij$ label unit cells, $mn$ label sublattices, $\alpha\beta$ label spin components, $\hat{d}_{im,jn}$ is a unit vector pointing from site $im$ to site $jn$, and $\bf \sigma$ is the vector formed by three Pauli matrices. As discussed above the spin-orbit coupling vector $\hat{\eta}_{mn}$ is chosen to be parallel to $\bm r^c_{mn} - (\bm r_m + \bm r_n)/2$, where $\bm r^c_{nm}$ is the mean of all neighbors of the bond $mn$. The band structure of this model is illustrated in Fig.~\ref{fig:1} (b). This $s$-$d$ model allows us to calculate ${\bf M}_{\rm orb}$ and $\overleftrightarrow \chi_{\rm os}^i$, but not the full $E_{\rm ani}$ that should come from a microscopic Hamiltonian of the $d$ electrons. We thus supplement the model with a phenomenological site-dependent uniaxial anisotropy of the exchange fields \cite{leblanc_2013} consistent with the crystal symmetry.

\begin{figure}
	\begin{center}
		\includegraphics[width=2.2 in]{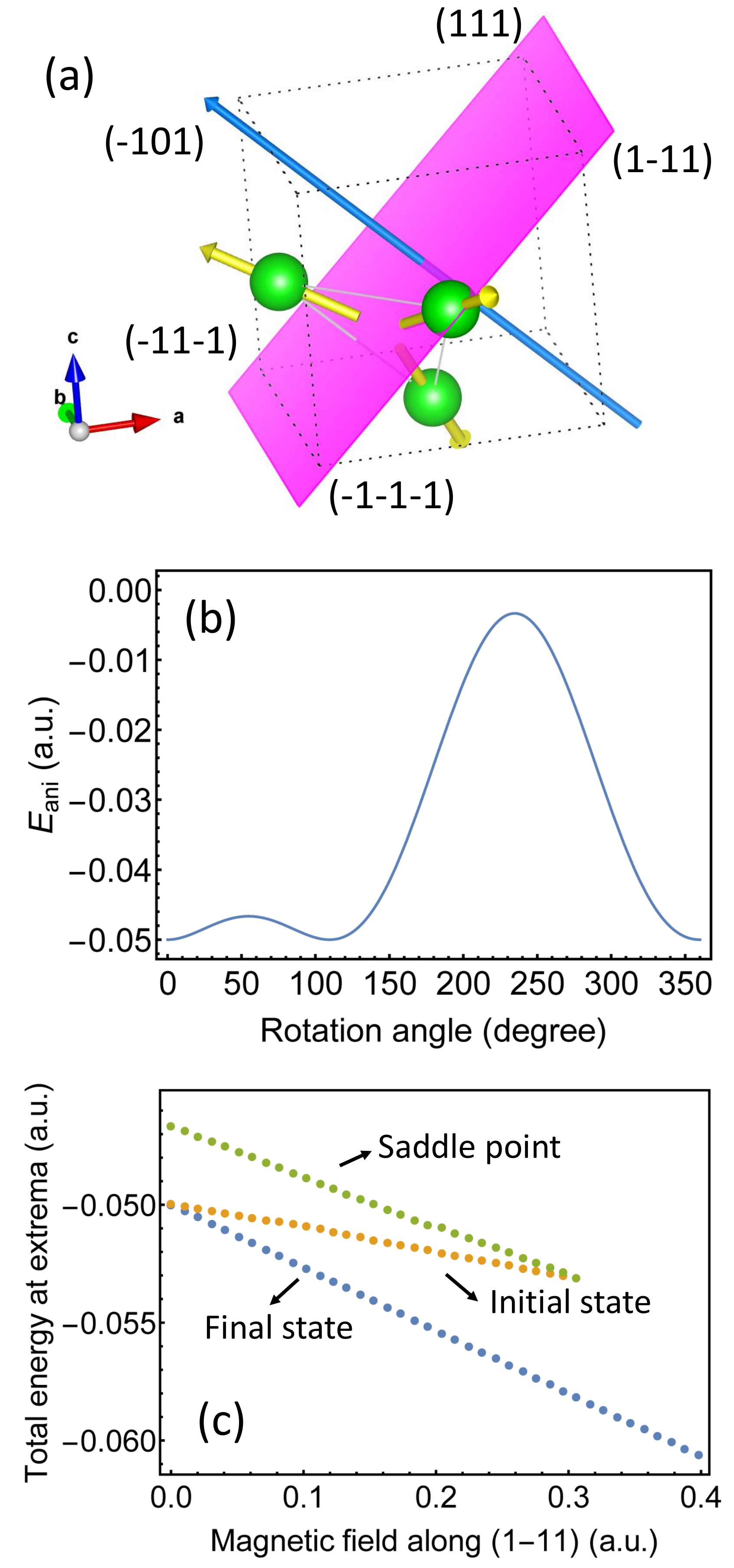}
	\end{center}
	\caption{(a) Rigid counterclockwise rotation of the noncollinear order parameter with respect to the  $(\bar{1}01)$ direction. The unrotated structure has orbital magnetization along (111). (b) Anisotropy energy vs. rotation angle. (c) Total energy at three lowest extrema along the rotation path vs. strength of an external magnetic field along $(1\bar{1}1)$. }\label{fig:2}
\end{figure}

Consider the starting ground state configuration with ${\bf M}_{\rm orb}$ along the (111) direction, and site-dependent exchange fields with 120$^\circ$ relative orientations in a perpendicular plane. The eight equivalent (111) directions have identical energy minima in the absence of a magnetic field. We apply a field $\bf H$ along the $(1\bar{1}1)$, with the expectation that with increasing $H$ the system will eventually switch to a configuration with a parallel ${\bf M}_{\rm orb}$. Based on symmetry considerations we focus on the path in SO(3) defined by rotation around the $(\bar{1}01)$ direction with angle $\theta$. If the order parameter were that of an ordinary ferromagnet, $E_{\rm ani}$ would, in the absence of a magnetic field, have four equivalent minima along this path at $\theta=0,\arccos(-1/3) \approx 109.47^\circ,180^\circ$ and $\arccos(-1/3)+180^\circ$ corresponding to four of the eight $(111)$ directions. However, plotting our $E_{\rm ani}$ vs. $\theta$ in Fig.~\ref{fig:2} (b) shows only two energy minima located at the first two rotation angles. The other two orientations differ in the chirality of the three exchange fields and do not have the same energy. Among the two remaining minima, $\theta=\arccos(-1/3)$ rotates the $(111)$ plane normal to the $(\bar{1}1\bar{1})$ direction. However, ${\bf M}_{\rm orb}$ is surprisingly rotated \emph{oppositely} to the $(1\bar{1}1)$ direction. (Similar behaviors exist in Mn$_3$Sn and Mn$_3$Ge \cite{nagamiya_1982}). Thus the magnetic switching induced by a field along $(1\bar{1}1)$ corresponds to reaching the minimum at $\theta=\arccos(-1/3)$ through the saddle point initially at $\theta\approx 55^\circ$. 

Fig. \ref{fig:2} (c) shows the energies of these three extrema as a function of $H$. As $H$ increases, the energy of the final $\theta=\arccos(-1/3)$ state moves below that of the initial minimum, and the latter eventually disappears after merging with the saddle point. At this time the magnetization configuration will switch to the final state $\theta=\arccos(-1/3)$. Switching between time-reversed states for noncollinear AHE AFMs is more complicated since it may not be able to be achieved through a single rotation around a fixed axis. One example is shown in \cite{supp}. In general one needs to consider the 3 degrees of freedom of SO(3), at least locally, in order to determine the smooth switching path connecting two time-reversed states. These points are also relevant to the formation and dynamics of magnetic domain walls driven by magnetic fields. 

{\it Discussion}---We have been ignoring spin contributions to the coupling with $\mathbf{H}$. For AHE AFMs with orbitally-dominant magnetization in the absence of a magnetic field, the orbital magnetization can still be shown to dominate in finite fields, by treating spin-orbit coupling as a perturbation. More generally, spin coupling to magnetic fields can be included in our formalism in a similar way as the orbital coupling, but through the spin-spin susceptibility that can also be obtained microscopically. Compared to the conventional method of using classical spin models combined with LLG equation in studying field-induced switching, our formalism has much fewer assumptions as all the essential quantities can be given by microscopic calculations. In particular, since $\overleftrightarrow{\chi}_{\rm os}$ is proportional to spin-orbit coupling, our work indicates that if one were going to describe orbitally-dominated AHE AFMs with phenomenological Heisenberg-type models, the effective g-tensor will be strongly dependent on order parameter direction. 

\begin{acknowledgments}
HC and AHM were supported by SHINES, an Energy Frontier Research Center funded by DOE BES under Grant No. SC0012670 and by the Welch Foundation Research Grant No. TBF1473. HC was also supported by the start-up funding from CSU. TCW and GYG were supported by Ministry of Science and Technology of the Republic of China (MOST 104-2112-M-002-002-MY3). DX was supported by DOE BES Grant No. DE-SC0012509. QN was supported by DOE (DE-FG03-02ER45958, Division of Materials Science and Engineering), NSF (EFMA- 1641101) and Welch Foundation (F-1255). The authors are grateful to Satoru Nakatsuji, Masaki Oshikawa, Yasuhiro Tada, and Junren Shi for helpful discussions. 
\end{acknowledgments}

\end{document}